\input harvmac
\noblackbox

\font\ticp=cmcsc10
 
\def\Title#1#2{\rightline{#1}\ifx\answ\bigans\nopagenumbers\pageno0\vskip1in
\else\pageno1\vskip.8in\fi \centerline{\titlefont #2}\vskip .5in}

\font\ticp=cmcsc10

\input epsf
\ifx\epsfbox\UnDeFiNeD\message{(NO epsf.tex, FIGURES WILL BE
IGNORED)}
\def\figin#1{\vskip2in}
\else\message{(FIGURES WILL BE INCLUDED)}\def\figin#1{#1}\fi
\def\ifig#1#2#3{\xdef#1{fig.~\the\figno}
\goodbreak\topinsert\figin{\centerline{#3}}%
\smallskip\centerline{\vbox{\baselineskip12pt
\advance\hsize by -1truein\noindent{\bf Fig.~\the\figno:} #2}}
\bigskip\endinsert\global\advance\figno by1}

%
%
\def\qn{quasinormal}

\def\om{\omega}

\def\[{\left [}
\def\]{\right ]}
\def\({\left (}
\def\){\right )}
\def\l{\ell}
\def\a{\rightarrow}


\lref\juan{J. Maldacena, Adv. Theor. Math. Phys. 2 (1998) 231, hep-th/9711200.}
\lref\schmidt{For a recent review, see
K. Kokkotas and B. Schmidt, ``Quasi-normal modes of stars and
black holes", gr-qc/9909058.}
\lref\chop{M. Choptuik, Phys. Rev. Lett. 70 (1993) 9.}
\lref\magoo{For a comprehensive review, see O. Aharony, S.S. Gubser,
J. Maldacena, H. Ooguri, and Y. Oz,
``Large N Field Theories, String Theory and Gravity",
hep-th/9905111.}
\lref\bdhm{T. Banks, M. Douglas, G. Horowitz, and E. Martinec,
``AdS Dynamics from Conformal Field Theory", hep-th/9808016.}
\lref\cama{C. Callan and J. Maldacena,  Nucl. Phys.
{\bf B472} (1996) 591,
hep-th/9602043.}
\lref\host{G. Horowitz and A. Strominger,  Phys. Rev. Lett.  77
(1996) 2368, hep-th/9602051.}
\lref\dama{S. Das and S. Mathur,  Nucl. Phys.  B478 (1996) 561,
hep-th/9606185.}
\lref\mast{J. Maldacena and A. Strominger, Phys. Rev. D55 (1997) 861,
hep-th/9609026.}
\lref\gkp{S. Gubser, I. Klebanov, and A. Peet, Phys. Rev. D54 (1996) 3915,
hep-th/9602135; A. Strominger, unpublished.}
\lref\stva{A. Strominger and C. Vafa, Phys. Lett.  B379 (1996) 99,
hep-th/9601029.}
\lref\hohu{G. Horowitz and V. Hubeny, ``Quasinormal Modes of AdS Black Holes
and the Approach to Thermal Equilibrium", hep-th/9909056.}
\lref\gund{For a review, see
C. Gundlach, Adv. Theor. Math. Phys. 2 (1998) 1, gr-qc/9712084.}
\lref\horstr{G. Horowitz and A. Strominger, Nucl. Phys. B360 (1991) 197.}
\lref\bhrev{For a review of some of these solutions, see
M. Duff, R. Khuri and J.  Lu, Phys. Rep. 259 (1995) 213, hep-th/9412184.} 
\lref\pol{J. Polchinski, Phys. Rev. Lett. 75 (1995) 4724, hep-th/9510017;
``TASI Lectures 
on D-Branes",  hep-th/9611050.}
\lref\dig{See, e.g., R. Dijkgraaf, talk at Strings '99.}
\lref\gkt{S. Gubser, I. Klebanov, and A. Tseytlin, Nucl. Phys. B534 (1998) 202,
hep-th/9805156.}
\lref\btz{M. Banados, C. Teitelboim, and J. Zanelli, Phys. Rev. Lett. 69 (1992)
1849.}
\lref\dewit{G. Cardoso, B. de Wit, and T. Mohaupt, ``Deviations from the Area
Law for Supersymmetric Black Holes", hep-th/9904005.}
\lref\hoit{G. Horowitz and N. Itzhaki, JHEP 9902 (1999) 010, hep-th/9901012.}
\lref\baro{V. Balasubramanian and S. Ross,
``Holographic Particle Detection",  hep-th/9906226.}
\lref\hohu{G. Horowitz and V. Hubeny, ``Quasinormal Modes of AdS Black Holes
and the Approach to Thermal Equilibrium", hep-th/9909056.}
\lref\hopo{L. Susskind, ``Some Speculations about Black Hole Entropy in String Theory", hep-th/9309145; G. Horowitz and J. Polchinski, Phys. Rev. D55 (1997)
6189, hep-th/9612146.}
\lref\thi{G. Horowitz and J. Polchinski, Phys. Rev. D57 (1998) 2557,
hep-th/9707170; T. Damour and G. Veneziano, ``Self-gravitating
fundamental strings and black-holes",  hep-th/9907030.}
\lref\grla{R.~Gregory and R.~Laflamme, Phys. Rev. Lett, 70 (1993) 2837,
hep-th/9301052.}

\baselineskip 16pt
\Title{\vbox{\baselineskip12pt
\line{\hfil hep-th/9910082} }}
{\vbox{
{\centerline{Comments on Black Holes in String Theory}}
}}
\centerline{\ticp Gary T. Horowitz\footnote{}{
 To appear in the proceedings of the Strings '99 conference, Potsdam, Germany,
 July 1999.}}
 \bigskip
 \centerline{\it Physics Department, University of California,
 Santa Barbara, CA 93106, USA}
 \vskip 2cm
 \centerline{\bf Abstract}
\bigskip
A very brief review is given of some of the developments leading to our current
understanding
of black holes in string theory. This is followed by a discussion
of two possible misconceptions in this subject -- one involving the stability
of small black holes and the other involving scale radius duality. Finally,
I describe some recent results concerning quasinormal modes of black holes
in anti de Sitter spacetime, and their implications for strongly coupled
conformal field theories (in various dimensions).

\Date{October, 1999}

\newsec{Review}
This talk is divided into three parts. The first is a brief
review of some of the key developments leading to our current understanding of 
black holes in string theory.
This part will be very elementary, and not assume
much knowledge of string theory.
Next, I will try to clear up two misconceptions that I had until recently,
and that I have seen in the literature. Finally, I will describe some
recent work about black holes in anti de Sitter spacetime, and their
implications for the approach to thermal equilibrium in 
strongly coupled conformal field theories.

Since supergravity is the low energy limit of string theory, the study
of black holes begins by finding solutions to this theory with horizons.
Actually, there are several supergravity theories that arise in string
theory, starting with the ten and eleven dimensional theories.
Since we are in higher dimensions,
there are extended black holes, or black $p$-branes.  The simplest
solutions are products of $R^p$ with the 
$D-p$ dimensional Schwarzschild solution (where $D=10$ or $11$),
but more interesting solutions carry
charge associated with a $p+2$ form. The rank is $p+2$ since the solution has
$p$ spatial dimensions along the brane. Adding one for time and one for the
radius in the transverse space, one finds that a sphere $S$ which surrounds the
brane must have dimension $D-(p+2)$.
The charge is then $Q\sim \int_S {}^*F_{p+2}$.
This charge can be nonzero
even though there are no fundamental sources in supergravity, 
since all you need is nontrivial spacetime topology. This is exactly
analogous the existence of charged black holes in Einstein-Maxwell theory
without charged matter.
The first charged black $p$-branes that were found almost ten years \horstr\ ago
assumed maximal symmetry, so all fields were a function of only one
radial variable. These solutions depended on two parameters
which were the mass $M$ (really mass per unit volume) and charge.
Solutions existed only when $M$ and $Q$ satisfied a certain inequality.
Since then, the number of black $p$-brane solutions has grown enormously
as people have learned how to add multiple charges, rotation, traveling 
waves, etc. \bhrev.

Unlike supergravity, string theory does have sources for many of these charges
called D-branes \pol. The charge to mass ratio of these D-branes is exactly
the same as the extremal limit of the black $p$-branes,
so the latter can be interpreted
as the gravitational field of the D-brane. At weak coupling, this
gravitational field goes to zero and the low energy
excitations of $N$ parallel D-branes are described by an $SU(N)$
gauge theory. The strongly coupled description of the same excited system
should be a nonextreme black $p$-brane. By comparing the weak and strong
coupling descriptions, one had 
the possibility of understanding
black hole entropy  by counting quantum states for the first time.

As an example, consider $N$ three-branes. To keep all quantities
finite, it is convenient to compactify the directions along the brane into
a three torus. Then, in the extreme limit, the area of the horizon goes to zero,
which agrees with the fact that at zero excitation energy, the only state
in the gauge theory is the ground state. To compare the entropies, we want
to add energy to the system. Equivalently, we can consider 
nonzero temperature $T$. The effective coupling is $gN$ where $g$ is the
string coupling constant.
 When $gN \ll 1$
the system is a weakly coupled $3+1$ dimensional
gauge theory at temperature $T$.
When $gN\gg 1$ one has a near extremal black three-brane at the same 
temperature. One can compare the entropies and find \gkp 
\eqn\threebr{S_{bh} = {3\over 4} S_{gauge}}
where $S_{bh}$ is the Bekenstein Hawking entropy of the black three-brane.
So the gauge theory has roughly the right number of degrees of freedom to 
explain the entropy of near extremal  black three-branes.
The fact that they are not exactly the
same was not a surprise. At the time this was first computed, it appeared that
one had two different descriptions of the same system which were valid for 
different ranges of the parameter $gN$. They appeared to have no overlapping
region of validity.

However, there were other situations where the entropies agreed exactly.
These were
obtained by looking at solutions with more than one charge. For example,
suppose four dimensions of space are compactified on a small $T^4$. We can
take $Q_5$ five-branes and wrap them around the compact dimensions to produce
an effective string in six dimensions. One can then 
add $Q_1$ one-branes to this string. When $g^2Q_1 Q_5\ll 1$,
the low energy excitations are described by a $1+1$ dimensional
conformal field theory \dig.
When $g^2Q_1 Q_5\gg 1$ the system is described by
a black string in six dimensions. If one now adds a small  amount of energy
and compares the entropies, one finds complete agreement (for 
large charges) \stva
\eqn\agree{S_{bh} = S_{cft}}

Why is this working? For the special case where the momentum along the
effective string is equal to the added energy,
i.e., one excites
only right moving modes, there is unbroken supersymmetry. The momentum along
the string is like another charge, and the black string remains extremal.
In this case, one can show that the number of supersymmetric states should
not depend on the coupling. But the entropy turns out to agree even when 
supersymmetry is broken, e.g., when
you excite equal amounts of left and right moving modes \refs{\cama,\host}.
Even more
importantly, the spectrum of Hawking radiation also agrees \refs{\dama,\mast}. 

The situation was clarified by Maldacena \juan\ who took a low energy limit 
which decoupled the excitations of the branes
from the excitations of the strings off the branes.
At strong coupling, this same limit corresponded to considering strings
moving very close to the horizon of the black $p$-brane. In the cases
of interest, the excitation of the branes is described by 
a conformal field theory (CFT), and
the near horizon
geometry of the extremal  black $p$-brane
is a product of anti de Sitter (AdS) space and a sphere.
For example, in the case of the three brane,
this geometry is $AdS_5 \times S^5$ where the 
radii of curvature are equal.
For the one-brane five-brane system, the near horizon geometry is
$AdS_3\times S^3 \times T^4$. Since the conformal field theory
is well defined even at strong coupling, we obtain two different
descriptions which are now
{\it valid for the same range of parameters}.
This lead Maldacena to his famous 
AdS/CFT correspondence.
If one adds energy
to the system, the spacetime is not exactly AdS, but still approaches it
asymptotically. So the correspondence says that
string theory in spacetimes which 
asymptotically approach AdS times a sphere is completely described by a 
conformal field theory. There is growing evidence in support of this
remarkable conjecture \magoo.

How does this explain the entropy results?
For the  case of the three-brane, it is easy to see from the field theory
side why the entropy might change between weak and strong coupling.
As you increase the coupling constant you add
potential energy to each state and increase its energy, so the number of
states for given total energy goes down.
Similarly, from the gravity side one can understand the change in entropy
as follows.
The near horizon geometry of the near extremal
 solution is a black hole in AdS. As you lower
the string coupling, the spacetime curvature increases in string units. This
results in stringy corrections to the geometry, and hence corrections to
the black hole
entropy.
In light of these effects, one would expect the weak coupling and
strong coupling results to be related in a complicated way. The fact that
they are related by a simple factor of $3/4$ is rather mysterious and
still not understood.

In contrast, for the one-brane five-brane system,
one has a $1+1$ dimensional CFT whose entropy depends only on the central 
charge. This can be computed exactly,
and is independent of the coupling constant. On the gravity side,
the near horizon geometry turns out to be the product of a three
dimensional BTZ black hole \btz\ and $S^3\times T^4$. This is locally a 
space of constant curvature and probably does not receive string
corrections as $g\a 0$ \gkt. There are other systems where the entropy 
can be computed exactly without supersymmetry,
including near extremal four dimensional
black holes. But as far as I know, in all such cases 
the corresponding field theory is a $1+1$ dimensional CFT and the
near horizon geometry is a space of constant curvature.
(When there is unbroken supersymmetry, the entropy can be reproduced
for a wider class of black holes, including
higher order corrections to the Bekenstein Hawking entropy \dewit.) 

In light of the AdS/CFT correspondence, we can begin to translate questions
about black hole physics into questions about field theory.
For example,
the formation of a large black hole in AdS is not an exotic process in the
CFT. It corresponds to the field theory evolution of a very special
high energy state into a typical (approximately thermal) state. More
importantly, the formation and evaporation of a small black hole in AdS
should be described by the usual unitary evolution in the field theory.

\newsec{Misconceptions}

We now come to our first possible
misconception, which involves small black holes in 
AdS. Let $r_+$ denote the horizon radius, and $R$ denote the radius of AdS.
The temperature
of a black hole in AdS decreases with mass for $r_+ \ll R$,
but 
increases with mass for $r_+\gg R$. So large black holes have
positive specific heat and are stable. Small black holes have negative 
specific heat and one often concludes that they are unstable and will
evaporate. However, let us compare the entropy of a small black hole
with the entropy of the same amount of energy in a thermal gas.
We will consider the case of black holes in $AdS_5\times S^5$. Since
the magnitude of the curvature on $S^5$ is the same as $AdS_5$, a gas
of radiation will be effectively ten dimensional. Since the curvature
of AdS acts like a confining box of side $R$, the
gas has $S\sim T^9R^9$ and $E \sim T^{10} R^9$, so
\eqn\entgas{S_{gas} \sim (RE)^{9/10}  }
A small black hole in $AdS_5$ which is uniform over the $S^5$
is unstable to localizing on the $S^5$
due to the Gregory-Laflamme instability \grla. So we should use ten
dimensional black holes which have
$S_{bh}\sim r_+^8/\l_p^8$ (where $\l_p$ is the ten dimensional Planck scale) 
and $E\sim r_+^7/\l_p^8$, which implies $S_{bh} \sim
E r_+$. Now let $R^8 \sim N^2 \l^8_p$, so $N$ is a measure of how
large the $S^5$ (or $AdS_5$) is in Planck units. (In the AdS/CFT correspondence,
this is the same $N$ that appears in the  group $SU(N)$, but since
we are asking a pure supergravity question, we don't need to introduce
any string theory or gauge theory quantities.)
So the entropies will be equal when
\eqn\equal{S_{bh} \sim {N^2 r_+^8\over R^8} \sim (RE)^{9/10} \sim 
\({N^2 r_+^7\over R^7} \)^{9/10}.}
This implies
\eqn\ans{ {r_+\over R} \sim {1\over N^{2/17}}  }
which can be made arbitrarily small for large $N$. In other words,
if $r_+/R > N^{-2/17}$, the black hole has more entropy than a 
gas in AdS. So its evaporation would violate the second law of thermodynamics.
What happens?

  If you fix the temperature, a small black hole with high temperature will
simply absorb energy from the heat bath until it turns into a large
black hole with the same temperature which is stable. But this is rather
unphysical since its hard to connect a heat bath to AdS, and this is not
the right boundary conditions when a black hole evaporates.
One should instead fix the total
energy, and consider a system consisting of both a black hole and
radiation.
It  is clear that if you start with all the energy in the black hole and
radiate a small amount $\epsilon$, $\delta S_{bh} \sim -\epsilon$ and
$\delta S_g \sim \epsilon^{9/10}$. So $\delta S_g + \delta S_{bh}>0$ and
you increase the
total entropy by starting to radiate. This is  a consequence of the negative
specific heat. But to see the final outcome, we must maximize the total 
entropy for given energy.
Let us divide the total energy into a part which is the gas, 
and a part which is the black hole: $E= E_g + E_{bh}$. As a crude approximation,
we will assume the entropy of the gas is the same that it would be in
the absence of the black hole. This may be justified since  we are
considering small black holes.
The total entropy is then
\eqn\total{ S\sim (E_gR)^{9/10} + E_{bh} (E_{bh} \l_p^8)^{1/7}  }
Using $\l_p^8 \sim R^8/N^2$, the second term becomes $(E_{bh}^8 R^8/N^2)^{1/7}$.
Extreming $S$ keeping the total energy fixed yields 
\eqn\extremum{ E_g^7 E_{bh}^{10} \sim {N^{20}\over R^{17}} }
Since the left hand side has a maximum when $E_g \sim E_{bh}$, we clearly
need $ER >N^{20/17}$ in order to have a stable equilibrium. (One can easily
check that this is equivalent to our earlier condition $r_+/R > N^{-2/17}$.)
When this condition is satisfied, there are two extrema of the entropy,
a local maximum when $E_{bh} > E_g$ and a local minimum when $E_{bh}< E_g$.
When $ER \gg N^{20/17}$,  the ratio $E_g/E_{bh}$ is very small in the 
maximum entropy configuration. 
So the net result is that if you fix the total energy,
most black holes will evaporate slightly and quickly come into equilibrium
with their Hawking radiation\foot{This is a more refined version of the
discussion in \bdhm.}. This is very similar to earlier studies of 
a black hole in a box.

A simple check on our result is the following.
In order for the black hole to be in stable equilibrium with the radiation,
it has to be large enough that it does not evaporate before the radiation
has a chance to see the background curvature. In other words, 
the lifetime of the black hole
must be larger than $R$. It is easy to check that in the stable regime,
all black holes
satisfy this condition. Ignoring the background curvature, 
the lifetime of a small ten dimensional black hole
can be computed from 
\eqn\lifetime{{dE\over dt} \sim T^{10} r_+^8 \sim {1\over r_+^2} }
Since $E \sim r_+^7/\l_p^8$, the lifetime is $t_0 \sim r_+^9/\l_p^8 \sim
N^2 r_+^9/R^8$. This will be of order $R$ when $r_+/R\sim N^{-2/9}$.
Since $N^{2/9} \gg N^{2/17}$ for large $N$,
this is much smaller than our lower bound for 
a black hole to be in stable equilibrium.
In the context of string theory, another thing we should check is whether
the size of the black hole remains bigger
than the string scale $\l_s$. Since
the smallest stable black hole has $r_+/R\sim
N^{-2/17}$, and $R^4 \sim gN \l_s^4$, we see that $r_+$ will be larger
than $\l_s$  provided $gN > N^{8/17}$. 

A similar calculation in eleven dimensions shows that black holes have more
entropy than a gas of radiation provided $r_+/R > (\l_p/R)^{9/19}$
where $\l_p$ is now the eleven dimensional Planck scale.

A second possible
misconception concerns the relation between AdS radius and scale
size in the CFT.  
In Poincare coordinates, the AdS metric is simply
\eqn\ads{ ds^2 = {r^2\over R^2} (-dt^2 + dx^idx_i) + {R^2\over r^2} dr^2}
Since this metric is 
invariant under $r\a \lambda r,\ (t,x^i) \a \lambda^{-1}(t,x^i)$,
it is often assumed that radial position in AdS is reflected in the
scale size of the corresponding excitation in the field theory. This
has been checked and is certainly true in many cases, usually involving
static configurations.
A particularly simple
example of this seemed to be a null particle moving radially in AdS.
It produces a gravitational shock wave which is reflected in the field
theory by a $<T_{\mu\nu}>$ which is concentrated on the null cone \hoit.
So the expanding excitation in the CFT is correlated with decreasing
radial position. However, we should ask what happens if the particle changes
its orbit inside AdS. The answer turns out to be that $<T_{\mu\nu}>$ continues
to grow at the speed of light even if the particle stops at some radius
$r$!

This is essentially a consequence of causality: $<T_{\mu\nu}>$ is determined
by the asymptotic form of the spacetime metric, and this metric 
is causally related to the
null particle. 
Letting $z=R^2/r$, \ads\ becomes
\eqn\adsz{ds^2 = {R^2\over z^2}(-dt^2 + dx_i dx^i + dz^2) }
 A particle initially falling in from $t=x_i=z=0$ can only influence
 fields inside its future light cone $t^2 \ge x_i x^i +z^2$. A point
 on the boundary $z=0, t=t^0,x_i = x_i^0$ can only be affected by
 events inside its past light cone $ (t^0 -t)^2 \ge (x_i-x_i^0)^2 + z^2 $.
 Looking at the intersection of these two sets, its clear that the
 maximum  $z$ value for the particle that can affect the asymptotic 
 field at $t^0,x_i^0$ is
 \eqn\zmax{z_{max} = {1\over 2} [(t^0)^2 - (x_i^0)^2]^{1/2}  }
 which occurs at $t=t^0/2$ and $x_i = x_i^0/2$. Therefore, as
 $(x_i^0)^2 \rightarrow (t^0)^2$, i.e., one
 approaches the light cone on the boundary,
 $z_{max} \rightarrow 0$. This means that even if the particle, or better
 yet, a rocket ship, stops
 at a constant value of $z$ inside AdS, the field will continue to grow along
 the light cone on the boundary.  Of course changing the trajectory inside
 will produce additional gravitational waves which will result in a change
 in the expectation value of the stress tensor {\it inside} the light cone.
 But the main lesson is that, in dynamical processes,
 the size of the disturbance
 on the boundary is NOT always a measure of the radial position of the particle
 in the interior. (For further examples of this phenomenon, see \baro.)

\newsec{Quasinormal modes}

A spherical black hole in $AdS_d$ is described by
\eqn\sadss{ds^2 = - f(r) \, dt^2 + f(r)^{-1} dr^2 + r^2 \,  d\Omega_{d-2}^2 }
where
\eqn\f{
   f(r) \equiv 1 + {r^2 \over R^2} - \({r_0 \over r}\)^{d-3}. }
The black hole horizon $r=r_+$ is at the largest zero of $f$.
A particle falling into this black hole will produce gravitational
waves. A rough estimate for when this radiation reaches infinity is
just the time it takes for the particle to fall from infinity to the vicinity
of the black hole. For a large black hole $r_+\gg R$, this is of order $1/T$
where $T\sim r_+/R^2$ is the black hole temperature.
At late times, this radiation  
is independent of the details of what fell in. It is described by characteristic
oscillations of the black hole geometry known as quasinormal modes \schmidt.
These oscillations are damped and the corresponding quasinormal frequencies
are complex. The mode with the smallest imaginary part dominates at late time
and gives the timescale for generic perturbations to decay. 
My student V. Hubeny and I have recently computed these
quasinormal frequencies. (For a more complete discussion,
see \hohu.) 

The damping time of these oscillations have important 
implications for the dual CFT. Suppose we start with a large static
black hole with temperature $T$. This is described in the field theory
by the thermal state\foot{For a black hole formed from
collapse of a pure state, the CFT state will still be pure, but resemble the
thermal state for macroscopic observations.} with temperature $T$.
Perturbing the black hole, corresponds to perturbing this thermal state,
and the timescale for the decay of the perturbation is the timescale 
for the return to thermal equilibrium. This dynamical timescale is
extremely difficult to compute directly, but can be done relatively
easily using the AdS/CFT correspondence. For simplicity, we considered
perturbations described by a real scalar field like the dilaton. 

Since a black hole in AdS has two dimensionful parameters $R$ and $r_+$,
it is not obvious how the \qn\ frequencies
$\om$  will scale as we change the size of the 
black hole. But for large black holes $r_+\gg R$,
it turns out that there is an extra symmetry which ensures that $\om$
will be proportional to the black hole temperature.

\ifig\ldlg{For large black holes, $\om_I$ is proportional to the temperature.
The top line is $d=4$, the middle line is $d=5$ and the bottom line is
$d=7$.}
{\epsfxsize=9.5cm \epsfysize=5.5cm \epsfbox{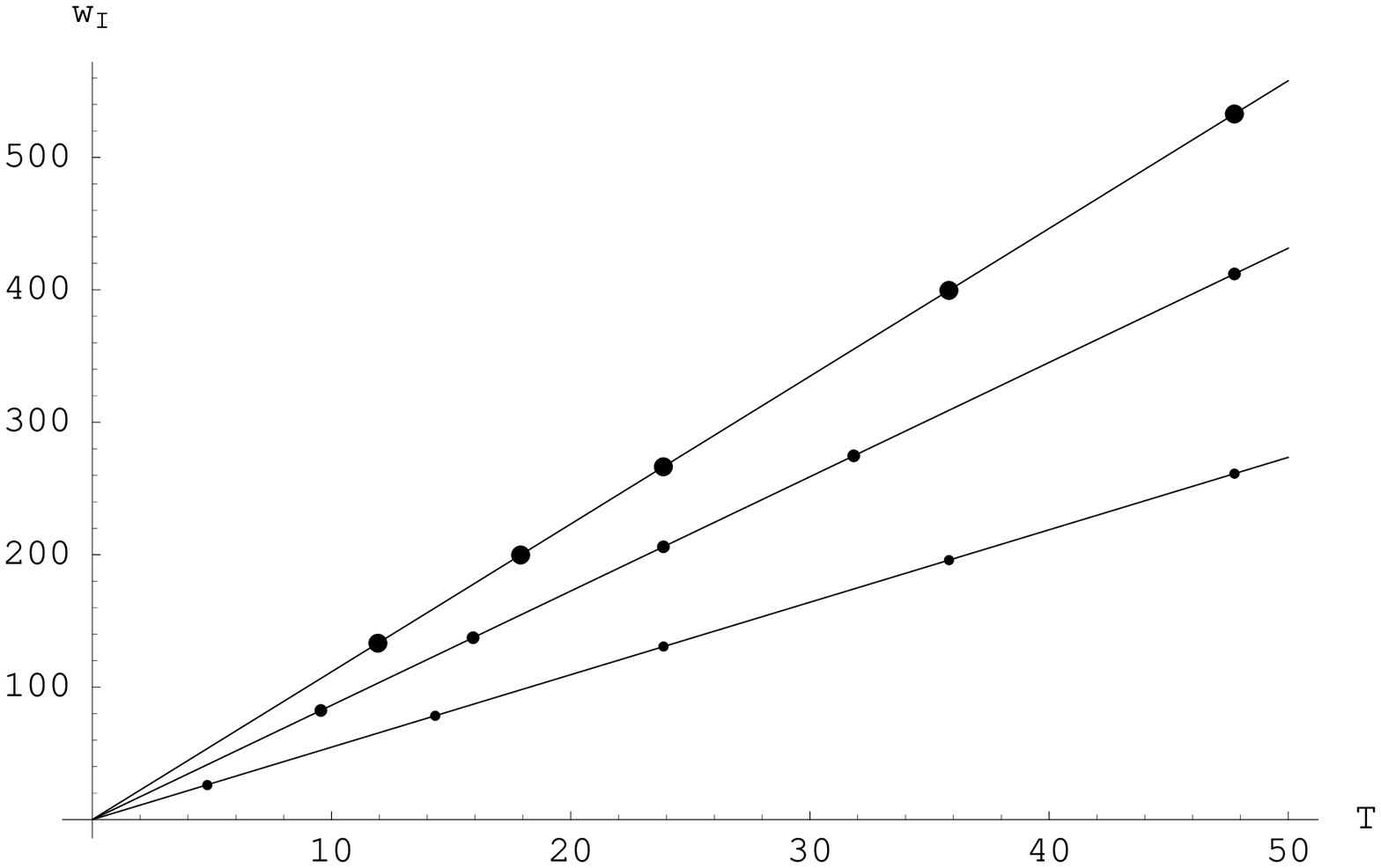}}

\ifig\lolg{For large black holes, $\om_R$ is also
proportional to the temperature.
The top line is now $d=7$, the middle line is $d=5$ and the bottom line is
$d=4$.}
{\epsfxsize=9.5cm \epsfysize=5.5cm \epsfbox{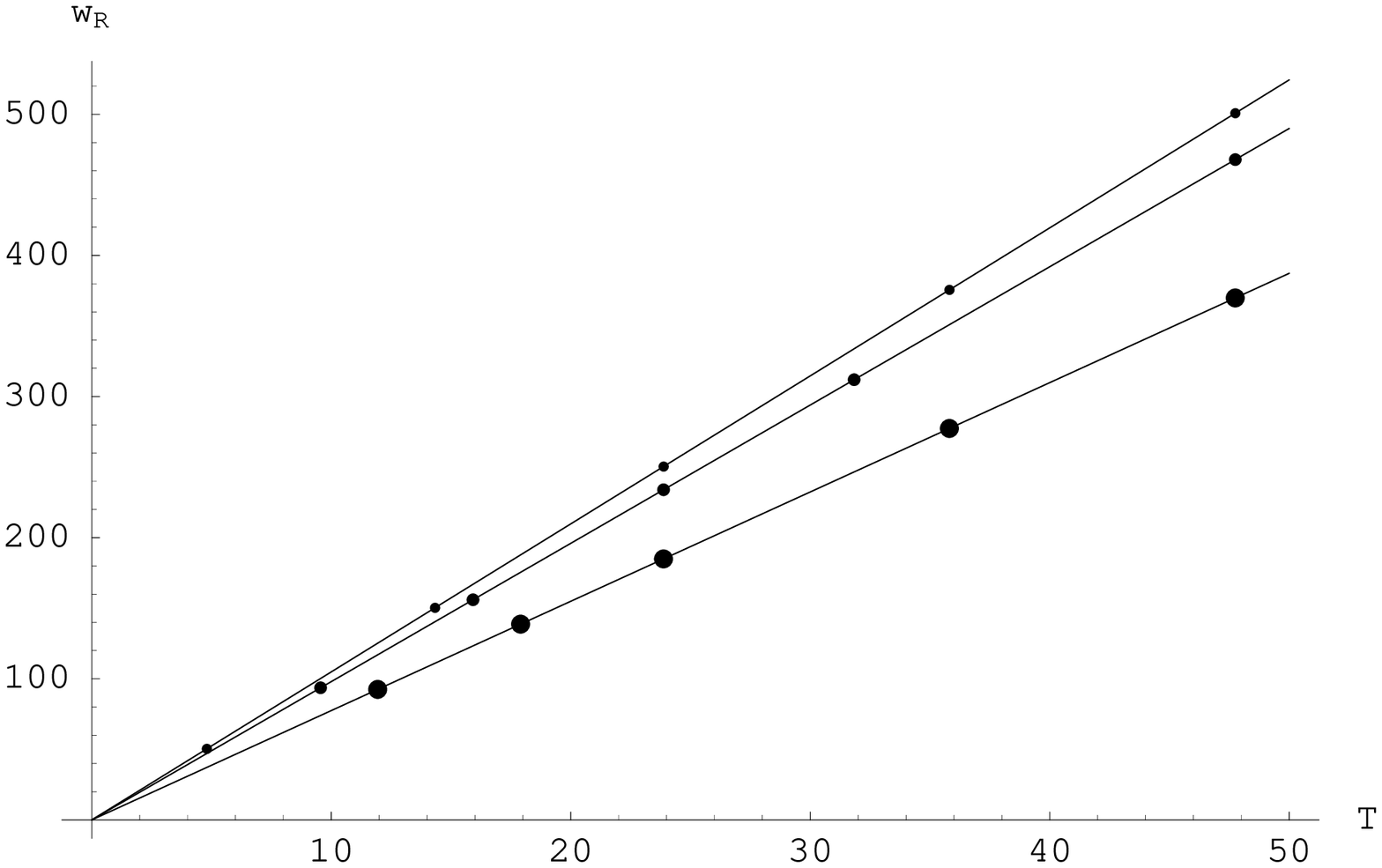}}

Let us decompose the quasinormal frequencies into real and imaginary parts
as $\om = \om_R - i\om_I$. (The sign is chosen so that exponentially decaying
modes correspond to $\om_I>0$.) The linear dependence with temperature is
clearly  shown in \ldlg\ and \lolg, where $\om_I$ and $\om_R$
respectively
are plotted  as a function of the temperature
for the four, five, and seven dimensional cases. We have set the AdS radius
equal to one, so all quantities are measured in units of the AdS radius.
The dots, representing the lowest
\qn\ mode for each black hole, lie on straight lines through the origin.
 In \ldlg, the top line corresponds to the $d=4$ case,
 the middle line is the $d=5$ case,
 and the bottom line is the  $d=7$ case.
 Explicitly, the lines are given by
 $$\om_I = 11.16 \ T \qquad {\rm for} \ d=4$$
 $$\om_I = 8.63 \ T \qquad {\rm for} \ d=5$$
 \eqn\qntemp{\om_I = 5.47 \ T \qquad {\rm for} \ d=7}

\ifig\fsdiv{$\om_I$ for smaller black holes in four dimensions.
The solid line is
$\om_I = 2.66\ r_+$, and the dashed line is $\om_I = 11.16\ T$.}
{\epsfxsize=9.5cm \epsfysize=5.5cm \epsfbox{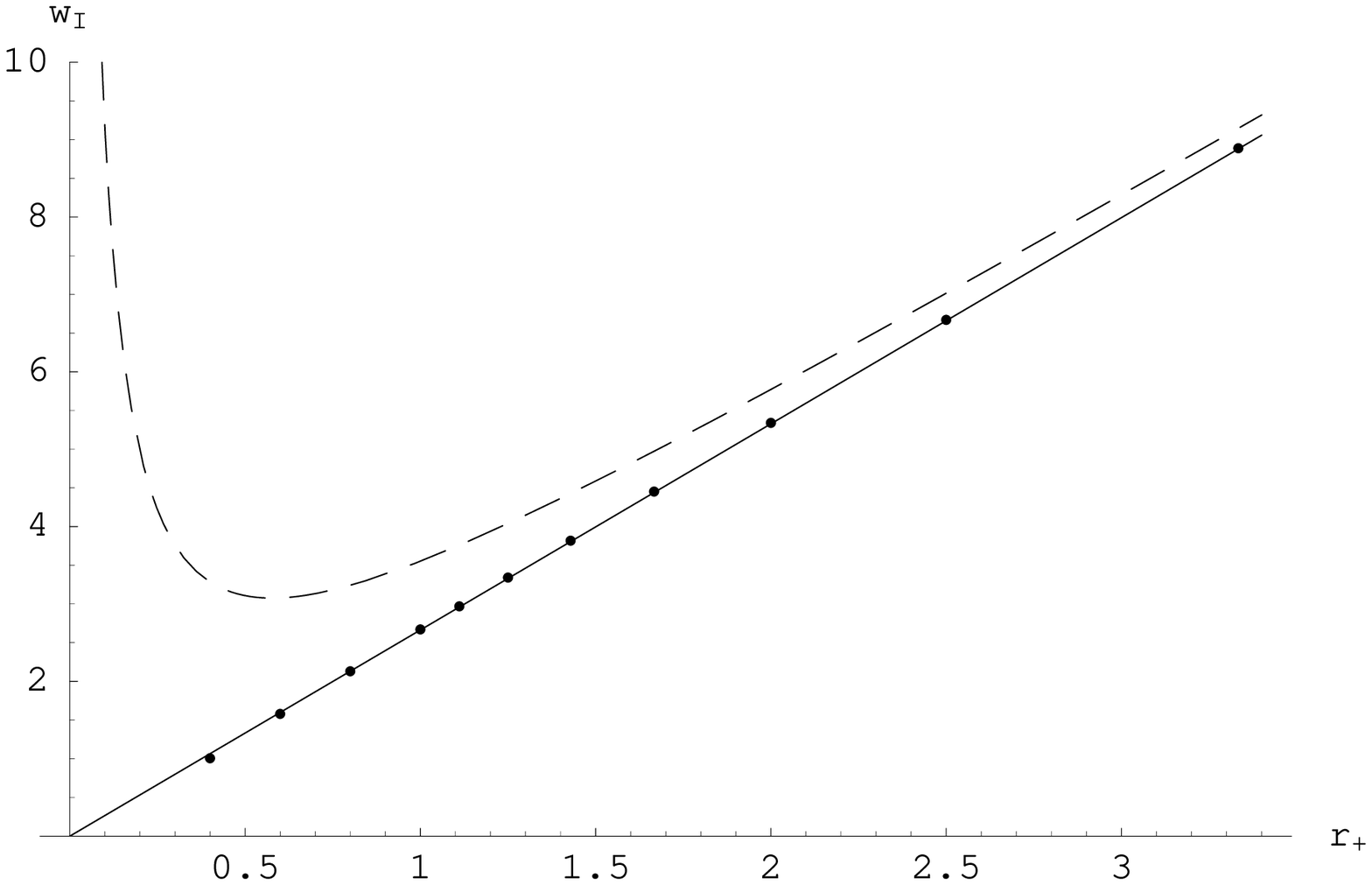}}

For smaller
black holes, 
the \qn\ frequencies do not scale with the temperature.
This is clearly shown in \fsdiv\ which plots $\om_I$ as a function
of $r_+$ for $d=4$ black holes with $r_+ \sim 1$. To a remarkable
accuracy, the points continue to lie along a  straight line
$\om_I = 2.66\ r_+$. The dashed curve represents the continuation of
the curve $\om_I = 11.16\ T$ shown in Fig. 1 to smaller values of
$r_+$. (For large $r_+$ these two
curves are identical.) It is not yet 
clear what the significance of this linear
relation is for the dual CFT. As we have seen, these black holes are
stable if one fixes the total energy, and thus correspond to a class of stable
states in the field theory. This linear relation is describing
the timescale for the decay of perturbations of these states.

The fact that the quasinormal frequencies do not follow the temperature
is very different from small black holes in asymptotically flat
spacetimes. In that case, there is only one scale $r_+$ in the problem
and the frequencies must go like $T\sim 1/r_+$. It is different in AdS
simply because the boundary conditions at infinity have been changed.
It should not be surprising that even for small black holes, the late time 
behavior of fields is different in AdS than in an asymptotically flat spacetime.

There is a striking similarity between the slope of the line in Fig. 3
and a number that has been computed in a completely different problem.
If you study the gravitational collapse of spherically symmetric scalar fields
(in four dimensional asymptotically flat spacetimes),
one finds that weak waves scatter and go off to infinity while strong waves
collapse to form black holes. On the boundary between these two possibilities,
there is initial data which collapses to form a `zero mass black hole', which
is really a naked singularity \chop. 
All such initial data approach the same solution,
called the critical solution, near the singularity. This critical solution is
known to have one unstable modes which grows like $e^{2.67t}$ \gund. This number
is very similar to the slope $2.66$ that we found above. Despite
the fact that both numbers characterize exponential behavior of spherically
symmetric scalar fields in four dimensions, further investigation has
failed to find any confirming indications of a connection between black holes
in AdS and black hole critical phenomena. It
appears at the moment to be  just a numerical coincidence.

\ifig\fsdv{$\om_I$ for smaller black holes in five dimensions.
The solid line is
$\om_I = 2.75\ r_+$, and the dashed line is $\om_I = 8.63\ T$.}
{\epsfxsize=9.5cm \epsfysize=5.5cm \epsfbox{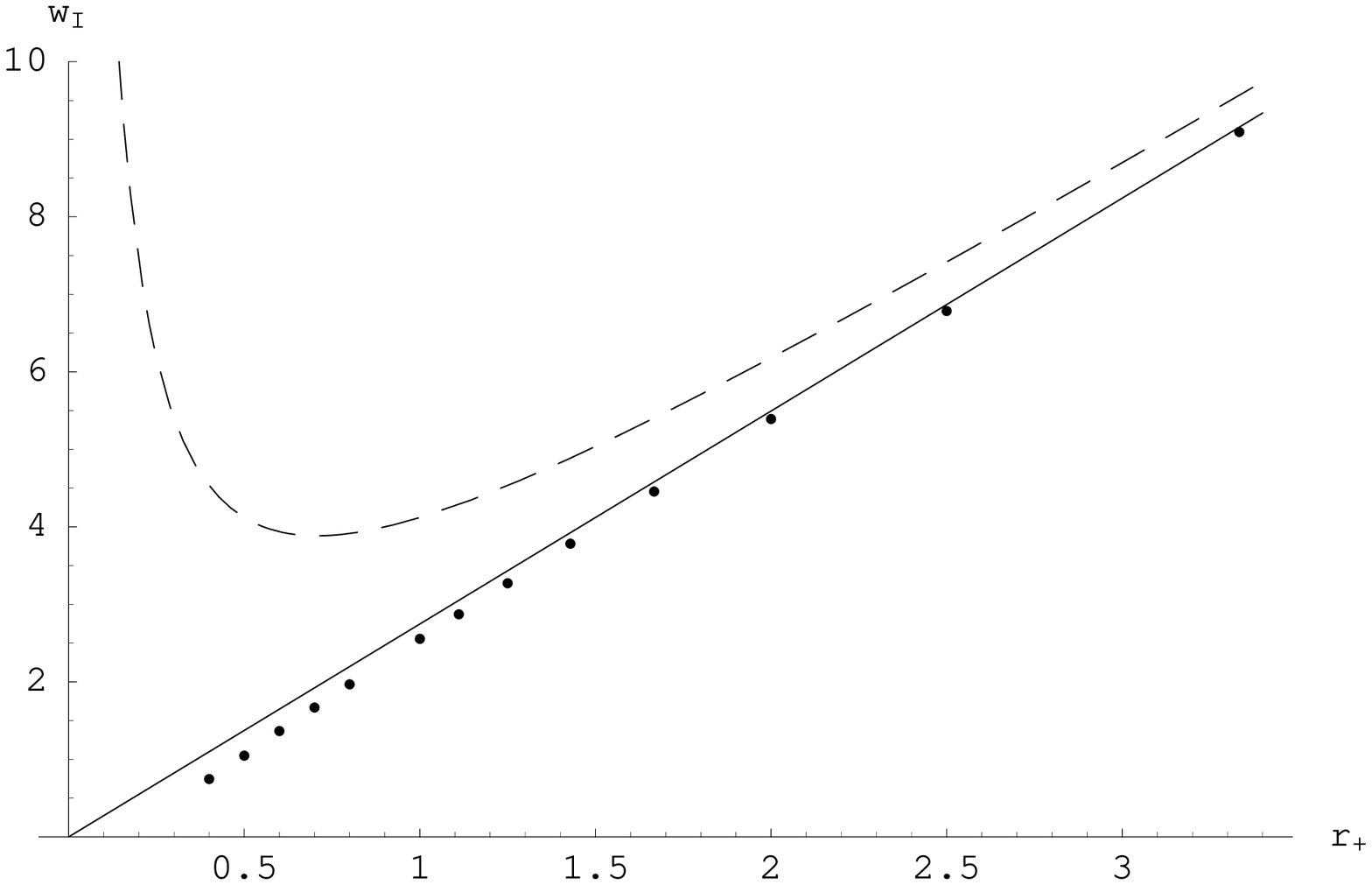}}

One reason for this is that the linear relation does not extend
to very small black holes. In fact,
since the \qn\ frequencies can be computed to an accuracy
much better than the size of the dots in Fig. 3, one can check that
the points
actually lie slightly off the line. This is  shown more clearly in
the five dimensional results in Fig. 4. Once again, the dashed curve
is the continuation of the curve $\om_I = 8.63\ T$ shown in Fig. 1,
and the solid curve is the line  $\om_I = 2.75\ r_+$ that it approaches
asymptotically.

\newsec{Conclusion}

If I was granted three wishes in the subject of black holes in string theory,
they would be:

\item{a)} Explain the $3/4$ factor relating the weak and strong coupling
calculations of the entropy of the near extremal
three-brane.

\item{b)} Find an exact calculation of the entropy of a Schwarzschild black hole.

\item{c)} Understand how (whether?)
the usual information loss arguments break down in the
evaporation of a small black hole.

We have already discussed (a). The current status of (b) is that
there are general arguments
which relate uncharged black holes to excited string states, and show that
the entropy should be proportional to the horizon area \refs{\hopo,\thi}.
But they are not yet
able to compute the numerical factor. Finally,
we mentioned
that in terms of the AdS/CFT correspondence, the evaporation of a small 
black hole in AdS should be a unitary process in the CFT. But we do not yet
understand how the usual semiclassical arguments for information loss break
down. This might point toward a possible limitation of the
AdS/CFT correspondence, but is more likely just a result of our current
lack of understanding of how the CFT describes the spacetime
inside the horizon. 

\vskip 1cm
\centerline{\bf Acknowledgements}
\vskip .5cm
I would like to thank the organizers of the Strings '99 conference
for a very stimulating meeting. This work was supported in part by
NSF grant PHY95-07065.

\listrefs
\end